\magnification=1200
\hsize =6.0 true in
\vsize = 8.0 true in
\hoffset=.375 true in
\voffset=.5true in

\font\mysmall=cmr8 at 8 pt

\def\f{\bf F}

\def\h{\bf H}

\def\z{\bf Z}

\def\p{\bf P}
\def\r{\bf R}
\def\dd{\cal D}

\def\ss{\cal S}

\def\picture #1 by #2 (#3){
		\vbox to #2{
		\hrule width #1 height 0pt depth 0pt
		\vfill
		\special {picture #3}}}
\font\teneufm eufm10 
\font\seveneufm eufm7 
\font\fiveeufm eufm5
\newfam\eufm
\textfont\eufm\teneufm
\scriptfont\eufm\seveneufm
\scriptscriptfont\eufm\fiveeufm

\noindent
\vskip 0.7 true in\noindent
\centerline {\bf  PATH INTEGRALS FOR A CLASS OF $P$-ADIC SCHR\"ODINGER
EQUATIONS}  \vskip
0.7 true in\noindent \centerline {V. S. VARADARAJAN}
\bigskip
\centerline{Department of Mathematics, University of California}
\centerline {Los Angeles, CA 90095--1555, USA} 
\centerline {\mysmall e-mail :  vsv@math.ucla.edu}
\vskip 1 true in\noindent
{\bf Abstract.\/} {\mysmall The theme of doing quantum mechanics on all abelian
groups goes back to Schwinger and Weyl. If the group is a vector space of
finite dimension over a non-archimedean locally compact division ring, it is of interest to examine
the structure of dynamical systems defined by Hamiltonians analogous to those encountered
over the field of real numbers. In this letter a path integral formula for the imaginary time
propagators of these Hamiltonians is derived.} \vskip 0.5 true in\noindent  {\bf AMS subject
classification (1991):\/}  81 S 40, 11 S 99.  \vskip 0.9 true in\noindent  {\bf 1. Introduction} 
\medskip\noindent General formulations of quantum theory when the configuration space is an
abelian group have been the theme of studies by Weyl$^1$ and Schwinger$^2$. The present
letter arose out of studies of quantum systems not only over the reals but over other fields and
rings, especially $p$-adic fields and adele rings$^{3,4,5}$. The main result is
that for an interesting class of Hamiltonians $H$ over nonarchimedean fields that are analogous to
the conventional ones, we can set up a formalism that leads to a path integral formula for the
propagators $e^{-tH} (t>0)$. The integrals are over the so-called {\it Skorokhod space\/}$^{6, 7}$ of
paths which allow discontinuities, but only of the first kind, namely that the left and right limits exist
at all time points and the paths are right continuous. This is to be contrasted with the real case
where the path integrals are with respect to conditional Wiener measures and so are on the space
of continuous paths.  \medskip  The study of quantum systems over finite and discrete structures
has been of interest for a long time$^8$. Interest in quantum structures over $p$-adic fields also
goes back a long way$^9$, but in recent years there has been quite a bit of activity, not only over
$p$-adic fields but over the adele rings of number fields also, and there are many treatments of 
these more general dynamical systems including path integral formulations$^{10,11}$. The
literature is extensive and an excellent review that includes a very good exposition of the basics of
$p$-adic theory is$^{12}$. But the  formalism presented in
this letter appears to be new.  \medskip
\medskip
I wish to thank S. R. S. Varadhan and Don Babbitt for some interesting discussions
on the ideas in this letter. 
 \bigskip \noindent 
{\bf 2. Vector spaces over local fields and division rings$^{13}$\/}  \bigskip\noindent   
We consider as configuration spaces vector
spaces over a division ring $D$ which is finite dimensional over a local (=locally compact,
nondiscrete, commutative) field $K$ of arbitrary characteristic. Unlike the case of the real field over
which there are only three division algebras of finite rank, namely, the reals, complexes, and the
quaternions, the theory of division algebras over nonarchimedean local fields is much richer and is
deeply arithmetical. Let $K$ be any nonarchimedean local field of arbitrary
characteristic and $D$ a division algebra of finite dimension over $K$. We shall  assume  that $K$
is the center of $D$; this is no loss of generality since we may always replace $K$ by the center of
$D$. Let $dx$ be a Haar measure on $D$ and $|\cdot|$ the usual  modulus function on $D$:
$d(ax)=|a| dx \  (a\not=0),\  |0|=0$. Then $|\cdot |$ is a multiplicative
norm which is  ultrametric (i.e., $|x+y|\le \max (|x|, |y|)$) that induces the original topology; and if we
define $R=\{x\in D | |x|\le 1\}, P=\{x\in D | |x| < 1\}$, then $R$ is the maximal compact subring of
$D$, $P$ is its maximal left, right, or two-sided ideal, $F:=R/P$ is a finite field of, say $q$,
elements, and there is an element $\pi $ such that $P=R\pi =\pi R$. Put $P^m=\pi ^mR=R\pi
^m (m\in {\z})$ and write $d_0x$ for the Haar measure for which $\int _R d_0x=1$. For any
nontrivial additive character $\chi $ of $D$ we write $\delta $ for the {\it conductor\/} of $\chi $; this
is the integer characterized by $\chi \big | _ {P^{-\delta }}\equiv 1,\quad \chi \big | _ {P^{-\delta -1
}}\not\equiv 1$. It follows from this that $$ \int _ {|x|\le q^m} \chi (x) d_0x=\cases { 0, & if $m\ge
\delta +1$ \cr q^m, & if $m\le \delta $\cr }\eqno(1) $$ \medskip
Let $W$ be a left vector space of finite dimension over
$D$. By a {\it $D$-norm\/} on $W$ is meant a function $|\cdot|$ from $W$ to the nonnegative reals
such that  (i) $|v|=0$ if and only if $v=0$ (ii) $|av|=|a||v|$ for $a\in
D$ and $v\in W$ (iii) $|\cdot|$ satisfies the ultrametric inequality, i.e.,  $ |u+v|\le \max
(|u|, |v|)\ (u, v\in W)$. The norm on the dual $W^\ast $ of $W$ 
is also a $D$-norm. If we identify $W$ and $W^\ast $ with $D^n$ by choosing dual bases,  and
define, for suitable constants $c_i>0$, $ |x|=\max _{1\le i\le n}(c_i|x_i|)\ (x=(x_1, x_2, \dots ,
x_n)\in W)$,  it is immediate that $|\cdot|$ is a $D$-norm on $W$ and $|\xi |=\max _{1\le i\le n}(c_i^{-1}|\xi
_i|)\ ( \xi  =(\xi _1,\dots ,\xi _n)\in W^\ast)$ defines the dual norm on $W^\ast $. It is known$^{13}$
that every $D$-norm is of this form. The set of values of $|\cdot|$ on $W\setminus (0)$ is  an
ordered set $$
0\dots <a_{-r}<a_{-r+1}<\dots <a_{-1}<a_0<a_1<\dots <a_s<\dots \eqno (2)
$$
where for some integer $m\ge 1$, $q^{{1\over m}}\le {a_{r+1}\over a_r}\le q\  (r\in
{\z})$ so that  $$
a_0q^{{r\over m}}\le a_r\le a_0q^r \qquad (r\ge 0)\qquad 
a_0q^r\le a_r\le a_0q^{{r\over m}} \qquad (r\le 0)\eqno (3)
$$ 
It is easy to see that there is a constant $A\ge 1$ such that 
$$
{1\over A}a^n\le  \hbox { meas }(\{ v | |v|\le a\}\le Aa^n \qquad \forall
a>0 \eqno (4)$$
A {\it $D$-lattice\/} in $W$ is a compact open $R$-submodule of $W$; these are the sets
of the form $\oplus _{1\le i\le n}Re_i \ ((e_i)_{1\le i\le n} \hbox { is a basis for } W)$. 
For any $u>0$, the set $\{v | |v|\le u\}$ is a $D$-lattice. For $x\in W, \xi \in W^\ast$, let $x\xi
$ be the value of $\xi $ at $x$. For any $D$-lattice $L$ in $W$ its {\it dual\/} lattice $L^\ast $is  the
set of all $\xi \in W^\ast $ such that $x\xi \in R$ for all $x\in L$. If $L$ is as above and $(\varepsilon
_i)_{1\le i\le n}$ is the basis of $W^\ast $ dual to $(e_i)_{1\le i\le n}$, then $L^\ast =\oplus _{1\le i\le
n} \varepsilon _iR$. If $L=\{x\ |\ |x|\le u\}$, then $$
\{x\ | \ |x|\le u^{-1}\} \subset L^\ast \subset \{x\ | \ |x|< qu^{-1}\} \eqno (5)
$$
Indeed, the first inclusion is clear from $|x\xi |\le |x||\xi |$. For the second, let $\xi \in L^\ast $
and choose $x_0\in W\setminus (0)$ such that $|x_0\xi |=|x_0||\xi |$. Replacing $x_0$ by $\pi
^rx_0$ for $r>>0$ we may assume that $x_0\in L$ and $\pi ^{-1}x_0\notin L$. Then $1\ge
|x_0||\xi |>q^{-1}u|\xi |$ so that $|\xi |<qu^{-1}$.     \medskip  Fix a nontrivial additive character $\chi
$ on $D$. Let ${\ss}(W)$ be the
Schwartz-Bruhat space of complex-valued locally constant functions with compact supports on
$W$. Let $dx$ be a Haar measure on $W$. Then ${\ss}(W)$ is dense in $L^2(W, dx)$, and the
Fourier transform ${\f}$ is an isomorphism of ${\ss}(W)$ with ${\ss}(W^\ast )$, defined by $$
{\f}(g)(\xi )=\int \chi (x\xi )g(x)dx\quad (\xi \in W^\ast )
$$ For a unique choice of Haar measure $d\xi $on $W^\ast $ (the {\it dual measure\/}) we have, 
$$
g(x)=\int \chi (-x\xi ){\f}g(\xi )d\xi \quad (x\in W, g\in {\ss}(W))
$$
If $W=D^n=W^\ast$ and $dx=q^{-n\delta /2}d_0x_1\dots d_0x_n$, then 
$d\xi =q^{-n\delta /2}d_0\xi _1\dots d_0\xi _n$ is the dual measure. 
\bigskip\noindent
{\bf 3. Hamiltonians over $W$}
\bigskip\noindent
Consider the $p$-adic Schr\"odinger theory which consists of the study of the
spectra of and semigroups generated by operators (``Hamiltonians") in $L^2(W)$ of the form
$H=H_0 + V$. Here $H_0$ is a pseudodifferential operator and $V$ is a multiplication operator.
Write $M_{W,b}$ for multiplication by $|x|^b
(b>0)$ in $ {\h}=L^2(W) $, and put
${\Delta}_{W,b}={\f}M_{W,b}{\f}^{-1}$. We consider Hamiltonians will of the form
$$
H_{W,b}={\Delta}_{W,b}+V
$$
Notice that for $D={\r}$ and $b=2$ this construction gives $H_{W,b}=-\Delta $. 
\bigskip\noindent 
{\bf 4. The probability densities $f_{t,b}$} \bigskip\noindent
We shall show that the dynamical semigroup $e^{-tH_{W,b}} (t>0)$
is just convolution by a one-parameter semigroup of {\it probability densities\/}. Lemma 2
contains the key calculation in the present letter and allows us to replace the Gaussian densities
of the conventional theory by these densities. $1_E$ is the characteristic function of the
set $E$. \bigskip\noindent
LEMMA 1. {\it Fix dual Haar measures $dx$ and $d\xi $ on $W$ and
$W^\ast $ respectively. Let $L$ be a $D$-lattice in $W^\ast $. Then
${\f}1_L=\hbox { meas } (L) 1_{\pi ^{-\delta }L^\ast }$. In particular,  $$ 
\int _L\chi (x\xi ) d\xi \ge 0\quad (x\in W)  $$
Moreover, if  $L=\{ \xi \ |\ |\xi |\le u\}$ where $u>0$, then
$$
\int _{|\xi |\le u}\chi (x\xi )d\xi =\cases {\hbox {meas } (L) & if $|x|\le q^\delta u^{-1}$\cr
0 & if $|x|\ge q^{1+\delta }u^{-1}$\cr}
$$}
\bigskip\noindent
{\it Proof\/}. This is standard. First assume that $W=D=W^\ast , L=R=L^\ast $. Then $d\xi = \hbox {
meas }(R) d_0\xi $. Now  $ \int _R \chi (x\xi )d_0\xi = |x|^{-1}\int _{|\zeta |\le
|x|} \chi (\zeta )d_0\zeta $ for $x\not=0$ so that
$$
\int _R \chi (x\xi )d_0\xi =1_{P^{-\delta }}(x)$$
from (2.1). The result for general $W, L$ is immediate since we may suppose that 
$W=D^n=W^\ast , L=R^n=L^\ast $. The last
assertion of the lemma follows from (2.5). \bigskip\noindent  
LEMMA 2. {\it Fix $t>0$ and $b>0$ and let $W$ be a $n$-dimensional left vector space
over $D$ with a $D$-norm $|\cdot|$. Then the function $\varphi $ on $W^\ast $ defined by  $$
\varphi (\xi )= \exp (-t|\xi |^b) \quad (\xi \in W^\ast ) $$   is in $L^m(W, d\xi )$ for all $m\ge 1$
and is the Fourier transform of a continuous 
probability density $f$ on $W$ with $f(ax)=f(x)$ for $x\in W, a\in D, |a|=1$. Moreover (i) $0<f(x)\le
f(0)\le A\ t^{-n/b}$ for all $t>0, x\in W$, $A$ being a constant $>0$ not depending on $t,x$ (ii) For
$0\le k<b$ we have, for all $t>0$ and a constant $A>0$ independent of $t$,  $$
 \int |x|^k f(x) dx \le A\ t^{k/b} 
$$}
\bigskip\noindent
{\it Proof\/}. From now on $A$ will denote a generic constant $>0$ independent of $t>0, x, \xi $. 
By (2.3), (2.4) we have, for $t>0$, 
$$
\int _{W^\ast } e^{-t|\xi |^b} d\xi =\sum _{r\in {\z}}e^{-ta_r^b}\int _{|\xi |=a_r} d\xi\le A\sum
_{r\in {\z}}a_r^n e^{-ta_r^b}<\infty  $$
Further 
$$\eqalign {
 \int _{W^\ast }e^{-t|\xi |^b} d\xi &=\sum _{r\in {\z}}e^{-ta_r^b}\int _{|\xi |=a_r} d\xi = \sum _{r\in
{\z}}e^{-ta_r^b}\int _{|\xi |\le a_r}d\xi -
 \sum _{r\in {\z}}e^{-ta_r^b}\int _{|\xi |\le a_{r-1}}d\xi \cr
&= \sum _{r\in {\z}}\left (e^{-ta_r^b}-e^{-ta_{r+1}^b}\right )\int _{|\xi |\le a_r}d\xi \le A\sum _{r\in
{\z}}a_r^n \left (e^{-ta_r^b}-e^{-ta_{r+1}^b}\right )\cr  
&\le A \sum _{r\in {\z}} t\int _{a_r^b}^{a_{r+1}^b} e^{-ty}y^{n/b} dy =At\int _0^\infty e^{-ty}y^{n/b} dy 
=At^{-n/b} \cr}
$$
So $\varphi \in L^m(W^\ast , d\xi )$ for $m\ge 1$. Set 
$$
f(x)=\int \chi (x\xi )e^{-t|\xi |^b}d\xi \quad (x\in W) 
$$
Clearly $f(ax)=f(x)$ for $|a|=1$. We prove that $f>0$ and $\in L^1(W, dx)$. As before, $$
f(x)=\sum _{r\in {\z}}e^{-ta_r^b}\int _{|\xi |=a_r}\chi (x\xi )d\xi 
=\sum _{r\in {\z}}\left (e^{-ta_r^b}-e^{-ta_{r+1}^b}\right ) \int _{|\xi |\le a_r}\chi (x\xi )d\xi \eqno
(1) $$
By Lemma 1, all the terms are $\ge 0$ and are $>0$ for $r<<0 $.
Hence $f(x)>0$ for all $x\in W$. Moreover Lemma 1 and (1) give 
$$
\eqalign {
\int |x|^k f(x) dx&=\sum _{r\in {\z}} 
\left (e^{-ta_r^b}-e^{-ta_{r+1}^b}\right ) \int _{|x|< q^{1+\delta }a_r^{-1}}|x|^kdx \int _{|\xi |\le a_r}\chi
(x\xi )d\xi \cr 
&\le A \sum _{r\in {\z}} 
\left (e^{-ta_r^b}-e^{-ta_{r+1}^b}\right ) a_r^{-k}\int _{|x|< q^{1+\delta }a_r^{-1}}dx \int _{|\xi |\le
a_r}d\xi \cr
&\le A \sum _{r\in {\z}} 
\left (e^{-ta_r^b}-e^{-ta_{r+1}^b}\right ) a_{r+1}^{-k} \le At\int _0^\infty e^{-ty}y^{-k/b} dy \le
At^{k/b}\cr}
$$
This proves in particular that $f\in L^1(W, dx)$ and completes the proof. 
\medskip 
Fix $b>0$ and write $f_{t,b}, \varphi _{t,b}$ for $f$ and $\varphi $. It is now clear that the
$(f_{t,b})_{t>0}$ form a continuous convolution semigroup of probability densities which goes to the
Dirac delta measure at $0$ when $t\to 0$. Hence for any $x\in W$ one can associate a $W$-valued
separable stochastic process with independent increments, $(X(t))_{t\ge 0}$, with $X(0)=x$, such
that for any $t>0, u\ge 0$, $X(t+u)-X(u)$ has the density $f_{t,b}$. As usual $E_x$ denotes the
expectation value with respect to this process. Clearly, when $b=2$ and $D={\r}$, this is the
Wiener process. \bigskip\noindent  {\bf 5. The paths of the stochastic processes $(X(t))_{t\ge 0}$
and $(X_{T,y}(t))_{t\ge 0}$} \medskip\noindent  Lemma 4.2 may be rewritten as follows.
\bigskip\noindent   LEMMA 1. {\it We have, for any $t>0$, $ E_0|X(t)|^k<\infty  (0\le k<b)$; and
for a fixed $k$, there is a constant $A_k>0$ such that $E_0|X(t)|^k\le A_kt^{k/ b}$ for all
$t>0$.} \bigskip
Let $D([0, \infty ) : M)$ be the space of right continuous functions on $[0, \infty )$ with
values in the complete separable metric space $M$ having only discontinuities of the first kind.
For any $T>0$ we write  $D([0,T] : M)$ for the analogous space of right continuous functions on
$[0,T)$ with values in the complete separable metric space $M$ having only discontinuities of the
first kind, and left continuous at $T$. These are the Skorokhod spaces$^{6,7}$ mentioned at the
beginning.
\bigskip\noindent  
LEMMA 2. {\it The process $X(t)_{t\ge 0}$ with $X(0)=x$
has paths in the space $D([0, \infty ) : W)$ and is concentrated in the subspace of paths
taking the value $x$ for $t=0$.}
\bigskip\noindent
{\it Proof\/}. It is immediate from the preceding proposition that for $0<t_1<t_2<t_3$,
$$
\eqalign {
E_x\{|X(t_2)-X(t_1)|^k|X(t_3)-X(t_2)|^k\}&=E_0\{|X(t_2)-X(t_1)|^k|X(t_3)-X(t_2)|^k\}\cr
&\le A (t_3-t_1)^{2k/b}\hskip 1.6 true in (1) \cr}
$$
So, if we take $k$ such that $b/2 < k < b$, we may use the
criterion of \v Centsov$^{14}$ to conclude the
required result. \medskip
We shall now construct the processes obtained from $(X(t))_{t\ge 0}$ by conditioning them to  go
through $y$ at time $t=T$. The density $f_{t,b}$ is everywhere positive and continuous and so the
finite dimensional conditional densities are defined everywhere and allow us to build the
conditioned process. We wish to prove that the corresponding probability measures can be defined
on the Skorokhod space $D([0, T] : W)$, and that they form a continuous family depending on the
starting point $x$ and the finishing point $y$. This will follow from the \v Centsov criteria in the
usual manner if we prove the following lemma.
\bigskip\noindent
LEMMA 3. {\it We have, uniformly for all $0<t_1<t_2<t_3<T$ with $|t_3-t_1|\le T/2$,
 and $z\in V$,  and for $b/2 < k < b$, 
$$
E_0\{ |X(t_2)-X(t_1)|^k |X(t_3)-X(t_2)|^k\ \big | X(T)=z\}\le A {1\over f_{T,b}(z)}(t_3-t_1)^{2k/b}
$$
}
\bigskip\noindent
{\it Proof\/}. The conditional expectation in question is (writing $f_t$ for $f_{t,b}$)
$$
\int |u_2|^k|u_3|^k
{f_{t_1}(u_1)f_{t_2-t_1}(u_2)f_{t_3-t_2}(u_3)f_{T-t_3}(z-u_1-u_2-u_3)\over f_T(z)}
du_1du_2du_3 
$$
Since $|t_3-t_1|\le T/2$, either $t_1$ or $|T-t_3|$ is $\ge T/4$, so that one of the two factors $
 f_{t_1}(u_1), \quad f_{T-t_3}(z-u_1-u_2-u_3)$ is bounded uniformly by a constant; the other factor
can then be integrated with respect to $u_1$ and the conditional expectation is majorized by
$$
A {1\over f_T(z)} E_0\{ |X(t_2)-X(t_1)|^k |X(t_3)-X(t_2)|^k\} 
$$
and the result follows from (1) and the  \v Centsov$^{14}$ criterion.
\medskip
The following theorem is now clear.
\bigskip\noindent
THEOREM 4. {\it There are unique families of probability measures ${\p}^b_x, {\p}^{T,b}_{x,y}
(x, y\in W)$  on $D([0, \infty ) : W)$ and $D([0, T] : W)$ respectively, continuous with respect to
$x$ and $(x,y)$ respectively, such that ${\p}^b_x$ is the probability measure of the $X$-process
starting from $x$ at time $t=0$, and ${\p}^{T,b}_{x,y}$ is the probability  measure for the
$X$-process that starts from $x$ at time $t=0$ and is conditioned to pass through $y$ at time
$t=T$.} \bigskip
It is now clear following the usual arguments$^{15}$ that one can obtain the
formula for the propagators. For simplicity let $V\ge 0$ and let the operator
$H_{W,b}$ be essentially self-adjoint on ${\ss}(W)$. This is the case if
$V$ is bounded, but see also$^{10}$.\bigskip\noindent {\bf Feynman--Kac propagator for
$e^{-tH_{W,b}} (t>0)$.\/} The operator $e^{-tH_{W,b}} (t>0)$ is an integral operator in $L^2(W)$
with kernel $K_{t,b}$ on $W\times W$  which is represented by the following integral on
the space ${\dd}_t=D([0, t] : W)$ : $$
K_{t,b}(x : y)=\int _{{\dd}_t} \exp \left ( -\int _0^t V(\omega (s))ds\right )dP^t_{x,y}(\omega
){\cdot} f_{t,b}(x-y) \quad (x, y\in W) $$
\bigskip\noindent
{\bf One dimensional case with $W=D$.\/}  Here 
$f_{t,b}(x)$ depends only on $|x|$ and so is known if we compute the values $f_{t,b}(\pi ^{-m})$.
We have, using the self-dual $dx=q^{-\delta /2} d_0x$,
$$
f_{t,b}(\pi ^{-m})=q^{-\delta /2}\sum _{r\le -m+\delta }q^r\left (e^{-tq^{rb}}-e^{-tq^{(r+1)b}}\right )
$$
\bigskip\noindent
{\bf  Coulomb problem.\/} Over $K$ one should take a three dimensional vector space. The
choice that is closest to what happens in the real case is the one where we take the {\it unique\/}
$4$-dimensional division algebra $D$ over $K$ and take for $W$ the subspace of elements of
$D$ of trace $0$ in the irreducible representation of $D$ in dimension $2$ over the
separable algebraic closure $K_s$ of $K$. Let the characteristic of $K$ be $\not=2$; then $D$ can
be described as a {\it quaternion algebra\/} generated by  ``spin matrices" and the
 analogy with the real case is really close. In fact$^{13,16}$ given any two elements $a,b\in
K^\times $ such that $b$ is not a norm of an element of
$K(\sqrt a)$ (such $a,b$ exist),  one can exhibit $D$ as the (``quaternion algebra") algebra
over $K$ with generators $i,j$ and relations $$
i^2=a, \quad j^2=b,\quad ij=-ji(=:k)
$$
One writes $(a,b)_K$ for this algebra and notes that $(-1, -1)_{\r}$ is just the usual algebra of
quaternions. Write $\sqrt a,
\sqrt {-b}$ for  square roots of $a,-b$ which are in $K_s$. If we define $$ \sigma _1=\pmatrix {0,
&\sqrt a\cr \sqrt a, &0\cr},\quad \sigma _2= \pmatrix {0,&-\sqrt {-b}\cr \sqrt {-b},&0\cr},\quad \sigma
_3= \pmatrix  {\sqrt a \sqrt {-b}, &0\cr 0, & -\sqrt a\sqrt {-b}\cr} $$
then $\sigma _1^2=aI,\quad  \sigma _2^2=bI, \quad \sigma _1\sigma _2=-\sigma _2\sigma
_1=\sigma _3$, and so there is a faithful irreducible representation $\rho $ of $(a,b)_K$ in
dimension $2$ such that $\rho (i)=\sigma _1, \rho (j)=\sigma _2, \rho (k)=\sigma _3$. Thus $D$ is
the algebra of matrices $$
x=\pmatrix { x_0+x_3\sqrt a \sqrt {-b}&x_1\sqrt a-x_2\sqrt {- b}\cr
x_1\sqrt a+x_2\sqrt {-b},&x_0-x_3\sqrt a \sqrt {-b}\cr}\qquad (x_j\in K)
$$Then  
$\det (x)=x_0^2-ax_1^2-bx_2^2+abx_3^2$, and $\det ^{1/2}$ is a $K$-norm on $D$. If we take
$W$ to be the subspace of $x\in D$ with  $Tr(x)=2x_0$ is $0$, one can study the Coulomb problem
on $W$. The Hamiltonian is  $$
H=\Delta _{D,b}-eM_{{1\over |x|}}\qquad (e>0 \hbox { a constant })
$$
where $M_{{1\over |x|}}$ is multiplication by ${1\over |x|}$. This is invariant under the group $U$ of
elements of determinant $1$ of $D$ which is {\it semisimple\/}. We shall treat these matters on a
later occasion.   \vskip 0.7 true in  \centerline {\bf
References\/} \bigskip
\item {1.} Weyl, H., {\it Theory of Groups and Quantum Mechanics\/}, Dover, 1931, Ch. III,
\S16, Ch. IV, \S\S 14, 15.
\smallskip   \item {2.} Schwinger, J., {\it Quantum Kinematics and Dynamics\/}, W. A.
Benjamin, 1970.
 \smallskip  
\item {3.} Digernes, T., Varadarajan, V. S., and Varadhan, S. R. S., {\it Rev. Math. Phys.\/} {\bf 6}
(1994), 621. 
\smallskip 
\item {4.} Varadarajan, V. S., {\it Lett. Math. Phys.\/} {\bf 34} (1995), 319.
\smallskip 
\item {5.} Digernes, T., and Husstad, E., and Varadarajan, V. S.,  {\it In preparation.\/}.
\smallskip
\item {6.} Skorokhod, A. V., {\it Dokl. Akad. Nauk. SSSR\/}, {\bf 104} (1955), 364; {\bf 106},
(1956), 781.   . \smallskip
\item {} Kolmogorov, A. N., {\it Theor. Prob. Appl.\/} {\bf 1},(1956), 215.  
\smallskip
\item {7.} Parthasarathy, K. R., {\it Probability measures on metric spaces\/}, Academic
Press, 1967. 
 \smallskip
\item {8} Stovicek, P., and Tolar, J., {\it Quantum mechanics in
a discrete space-time\/} Rep. Math. Phys. {\bf 20} (1984), 157.    
\smallskip
\item {} Beltrametti, E. G. {\it Can a finite geometry
describe the physical space-time?\/}, Atti del convegno di geometria combinatoria e sue
applicazioni, Perugia 1971.
\smallskip
\item {9.} Ulam, S. {\it Sets,
Numbers, and Universes, Selected Works\/}, MIT Press 1974. See paper [86] (p 265) with
commentary by E. Beltrametti, p 687.  
\smallskip
\item {10.} Vladimirov, V. S., and Volovich, I., {\it Lett. Math. Phys.\/} {\bf 18} (1989), 43  
\smallskip
\item {} Vladimirov, V. S., {\it Leningrad Math. J\/} {\bf 2} (1991), 1261.
\item {11.} Parisi, G., {\it Modern Phys. Lett. A3\/} (1988), 639
\smallskip
\item {} Meurice, Y., {\it Phys. Lett. B\/} {\bf
104} (1990), 245.\smallskip
\item {}  Zelenov, E. I., {\it J. Math. Phys.\/} {\bf 32} (1991), 147.
\smallskip   
\item {12.} Brekke, L., and Freund, P. G. O., {\it Physics Reports\/} {\bf 233} (1993), 1. 
\smallskip\item {13.} Weil, A., {\it Basic Number Theory\/}, Springer, 1967.
\smallskip 
\item {14.} \v Centsov, N. N., {\it Theor. Prob. Appl.\/} {\bf 1}, (1956), 140.
\smallskip 
\item {15.} Simon, B., {\it Functional integration and quantum physics\/}, Academic Press,
1979.
\smallskip
\item {16.} Shafarevitch, I. R., {\it Algebra \/}, Encyclopedia of Mathematical Sciences,
Vol 11, Springer Verlag, 1990.

\bye